\documentclass[aps,prd,final,twocolumn,floats,floatfix,nofootinbib,10pt]{revtex4-1}

\usepackage{blindtext}  
\usepackage{graphicx}
\usepackage{amsmath}
\usepackage{amsfonts}
\usepackage{amssymb}
\usepackage{amstext}
\usepackage{braket}
\usepackage[english]{babel}
\usepackage{helvet}
\usepackage{microtype}
\usepackage{listings}
\usepackage{mathtools}
\usepackage[pdftex]{hyperref}

\input{qcircuit.sty}

\begin{document}

\title{Decomposition of unitary matrix into quantum gates}

\author{Dmytro Fedoriaka}

\date{June 19, 2019}

\begin{abstract}
An algorithm is proposed to convert arbitrary unitary matrix to a sequence of X gates and fully controlled $R_y, R_z$ and $R_1$ gates. This algorithm is used to generate Q\# implementation for arbitrary unitary matrix. Some optimizations are considered and complexity of the result is analyzed.
\end{abstract}

\maketitle

\section{Introduction}

In this paper I will solve a problem of implementing a unitary matrix with a sequence of quantum gates which can be expressed using standard library of Q\# language. 

Q\# is a domain-specific programming language used for expressing quantum algorithms, developed by Microsoft \cite{qsharpDocs}. Its standard library currently doesn't support explicitly specifying unitary operation by a matrix. Instead, programmer has to express it as a sequence of built-in quantum gates. 
However, while designing some quantum algorithms it may be necessary to implement a unitary operation which is given by a matrix and decomposition of this matrix into standard gates is not obvious (for example, \cite[problem B2]{contestEditorial}).

In this paper I will describe algorithm which can be used to generate Q\# code using only fully-controlled $R_x$, $R_y$ and $R_1$ gates and single-qubit $X$ gates. Length of this code (in terms of number of commands) will be $\mathcal{O}(4^n)$, where $n$ is number of qubits.

\smallskip

I will start by giving some basic definitions. Then I will describe proposed algorithm (which is based on \cite{nakahara2008quantum, li2013decomposition}). Then I will analyze complexity of this algorithm. I will conclude with discussion of several related topics. 

\section{Definitions}

\textbf{Qubits.} \textit{Qubit} is a quantum system which can be in superposition of two basis states $\ket{0}$ and $\ket{1}$. 

\textit{Register} of $n$ qubits is a quantum system which consists of $n$ qubits and its state space is tensor product of state spaces of those qubits. Register's state space is span of $2^n$ basis states, each of which is tensor product of qubits' basis states (although not any register's state is a tensor product of qubits' states).

We denote states in qubit register of $n$ qubits by binary string consisting of $n$ bits. The leftmost bit in string corresponds to qubit $0$, rightmost bit corresponds to qubit $n-1$.
Also we denote state by an integer $i \in 0, \dots 2^n-1$, represented by that binary string in little-endian style (i.e. leftmost bit is least significant). For example, if $n=5$ then register has 32 basis states, and state 25 is

\begin{equation}
\ket{25} = \ket{10011} = \ket{1} \otimes \ket{0}
\otimes \ket{0} \otimes \ket{1} \otimes \ket{1}.
\end{equation}

If $i$ is index of state, $i[j]$ is $j$-th bit of this index (i.e. $j$-th character in binary string representing $i$). 

All notation is 0-indexed.

\textit{Quantum gate} acting on one qubit is a unitary operator acting on state space of this qubit. Similarly, quantum gate acting on register of several qubits is a unitary operator acting on space which is tensor product of state spaces of those qubits.

\bigskip

\textbf{Matrices.} Complex-valued matrix $A \in \mathbb{C}^{n \times n}$ is called \textit{unitary} if  $U^\dag = U^{-1}$.
$\mathrm{U}(n)$ is set of all unitary matrices of size $n \times n$. 

Matrix is called \textit{special unitary}, if it is unitary and has determinant 1. $\mathrm{SU}(n)$ is set of all special unitary matrices of size $n \times n$. 

\textit{Two-level unitary} matrix is a unitary matrix obtained from identity matrix by changing a $2\times 2$ principal submatrix. 

Any quantum gate on register of $n$ qubits, being an unitary operator, can be completely defined by unitary matrix $2^n \times 2^n$. Indexing of matrix elements follows the same pattern as indexing of register states, i.e.
$U_{ij} = \bra{i} U \ket{j}$, where $i,j \in [0, 2^n-1]$.

\bigskip

\textbf{Controlled gates}. Let's consider a gate $U$ acting on register $t$. Let's add a new qubit $c$ to this register and define new gate $CU$ as follows. If $c$ is in state $\ket{1}$, this gate applies $U$ on $t$, but if $c$ is in state $\ket{0}$, this gate doesn't change the register's state:
\begin{equation}
\begin{cases}
CU (\ket{a} \otimes \ket{0}) = \ket{a} \otimes \ket{0},\\
CU (\ket{a} \otimes \ket{1}) = (U\ket{a}) \otimes \ket{1}.
\end{cases}
\end{equation}

Such gate is called \textit{controlled}, qubits in $t$ are called \textit{target} qubits, and qubit $c$ is called \textit{control} qubit.

Example of a controlled gate is CNOT (controlled-X) gate:

\begin{equation}
CNOT = \begin{psmallmatrix}
1 & 0 & 0 & 0 \\
0 & 1 & 0 & 0 \\
0 & 0 & 0 & 1 \\
0 & 0 & 1 & 0
\end{psmallmatrix}.
\end{equation}

Similarly, let's define controlled gate with multiple control qubits. Gate $CU$ acting on register
$t_0, \dots, t_{T-1}, c_0, \dots, c_{C-1}$ is controlled by qubits $c_0, \dots, c_{C-1}$, if

\begin{equation}
CU (\ket{a} \otimes \ket{b}) =
\begin{cases} 
(U \ket{a}) \otimes \ket{b} & \text{if} ~\ket{b} = \ket {1 \dots 1}, \\
\ket{a} \otimes \ket{b} & \text{otherwise},
\end{cases}
\end{equation}

where $\ket{a}$ - basis state of target qubits
$t_0, \dots t_{T-1}$, $\ket{b}$ - basis state of control qubits $c_0, \dots c_{C-1}$, $U \in \mathrm{U}(2^T)$. In other words, it applies $U$ on target bits only if all control qubits are set to $\ket{1}$.

Matrix of a controlled gate has special form. It's an identity matrix $2^{C+T} \times 2^{C+T}$, where lower right submatrix $2^T \times 2^T$ is replaced by $U$. If $T=1$, this matrix is two-level unitary.

\bigskip
We will denote $I$ --- unitary matrix; $X = \begin{psmallmatrix}
0 & 1 \\ 1 & 0
\end{psmallmatrix}$ --- Pauli X matrix, also known as X gate or NOT gate.

Expression $a \oplus b$ means bitwise addition modulo 2 (also known as XOR), e.g. $25 \oplus 13 = 20$.

\section{Algorithm}

\subsection{Two-level decomposition}
First step is to decompose our unitary matrix $A \in \mathrm{U}(d)$ into product of two-level unitary matrices ($d=2^n$). Following algorithm is based on algorithm in  \cite{li2013decomposition}.

Let's make elements in the first column equal to zeroes by multiplying matrix (from the right) by two-level unitary matrices. Assume that at the current step elements $A_{0,i+1} \dots A_{0, d-1}$ are already zeroes and we want to make element $A_{0,i}$ zero as well, without affecting already eliminated elements. This can be written as:

\begin{equation}
\begin{psmallmatrix}
    \dots & a & b & 0 & \dots & 0 \\
    \dots & \dots & \dots & \dots &\dots & \dots 
\end{psmallmatrix} ~ U = 
\begin{psmallmatrix}
    \dots & c & 0 & 0 & \dots & 0 \\
    \dots & \dots & \dots & \dots &\dots & \dots 
\end{psmallmatrix}.
\end{equation}

Suppose $a \ne 0, b \ne 0$.

Matrix $U$ can be chosen to be two-level unitary matrix acting on elements $(i-1, i)$ with non-trivial unitary $2 \times 2$ submatrix $U'$, where

\begin{equation}
\label{cond1}
(a ~ b )~
U' = 
( c ~ 0 ).
\end{equation}

Let's show that we can always find such special unitary matrix $U'$ which satisfies condition \eqref{cond1}, and makes $c$ real positive number.

Any special unitary matrix $U'$ can be written in form  \cite[\S 4.6]{nakahara2008quantum}:

\begin{equation}
\label{su_decomp}
U' = \begin{pmatrix}
    \cos{\theta}e^{i \lambda}  & \sin{\theta}e^{i \mu} \\
    -\sin{\theta}e^{i \mu}  & \cos{\theta}e^{-i \lambda}
\end{pmatrix},
\end{equation}

where $\theta, \lambda, \mu \in \mathbb{R}$.

Substituting \eqref{su_decomp} in \eqref{cond1}, we get:

\begin{equation}
\label{cond_system}
\begin{cases}
a \cos \theta e^{i\lambda} - b \sin \theta e^{i \mu} = c, \\
a \sin \theta e^{i\mu} + b \cos \theta e^{- i \lambda} = 0.
\end{cases}
\end{equation}

From second equation we get:

\begin{equation}
\label{tan_theta}
\tan\theta = -\frac{b}{a} \exp \left(-i(\lambda + \mu) \right).
\end{equation}

Let's demand that $\tan \theta$ is real and positive. 
Then:
\begin{equation}
\label{theta_def}
\theta = \arctan\left(\left| \frac{b}{a} \right| \right),
\end{equation}

\begin{equation}
\label{mu_def_1}
\arg \left(-\frac{b}{a} exp(-i(\lambda + \mu)) \right) = \pi + \arg(b) - \arg(a) - \lambda - \mu = 0 
\end{equation}

From \eqref{mu_def_1} we can express $\mu$:

\begin{equation}
\label{mu_def}
 \mu = \pi + \arg(b) - \arg(a) - \lambda.
\end{equation}

Let's find $\lambda$. For this, let's express $c$ from first equation in \eqref{cond_system}, using \eqref{tan_theta}:

\begin{equation}
c = \cos\theta (a e^{i \lambda} - b \tan\theta e^{-i\mu}) = 
\cos \theta \left(a e^{i \lambda} + \frac{b^2 e^{-2 i \mu}}{a e^{-i \mu}} \right)
\end{equation}

Let $\lambda = - \arg a$. 
Then $a e^{i \lambda} = |a| e^{i \arg a} e^{- i \arg a} = |a|$.
From \eqref{mu_def} we get $\mu = \pi + \arg b$, therefore
$b e^{-i \mu} = |b| e^{i \arg b} e^{-i \pi} e^{-i \arg b} = -|b|$, so

\begin{equation}
c = \cos\theta \cdot \left(|a| + \frac{|b|^2}{|a|} \right).
\end{equation}

We have $c$ real and positive, exactly what we wanted, so matrix $U'$ is given by formula \eqref{su_decomp} with

\begin{equation}
\theta = \arctan\left(\left| \frac{b}{a} \right| \right);~ \lambda = - \arg(a); ~\mu = \pi + \arg(b).
\end{equation}

If $b=0$, we can just skip this step, formally putting $U' = I$. If $a=0$, we can just replace columns by taking $U' = X = \begin{psmallmatrix}
    0 & 1 \\ 1 & 0 
\end{psmallmatrix}$ and proceed.

After we finished eliminating first row, all elements in it except first will be zeroes and matrix still will be unitary. First element then must have magnitude 1 (because norm of row in unitary matrix must be 1). As our construction always makes $c$ real and positive, it must have value 1. All other elements in first column must be zeroes, because norm of first column must be 1. So, we get $(d-1)$-level matrix, and we can apply the same algorithm to remaining $(d-1) \times (d-1)$ submatrix and repeat it until only $2 \times 2$ non-trivial submatrix is left in $A$, which will make $A$ 2-level unitary matrix. 

For example, for matrix $4 \times 4$ this process looks like this:

\begin{equation}
\begin{pmatrix}
    *  & * & * & * \\
    *  & * & * & * \\
    *  & * & * & * \\
    *  & * & * & * 
\end{pmatrix}
\to
\begin{pmatrix}
    1  & 0 & 0 & 0 \\
    0  & * & * & * \\
    0  & * & * & * \\
    0  & * & * & * 
\end{pmatrix}
\to
\begin{pmatrix}
    1  & 0 & 0 & 0 \\
    0  & 1 & 0 & 0 \\
    0  & 0 & * & * \\
    0  & 0 & * & * 
\end{pmatrix}
\end{equation}

Let's denote $U_1$, $U_2$, \dots $U_D$ --- all matrices which we applied during this algorithm and $U_f$ - final two-level matrix we got. Then the whole process can be written as

\begin{equation}
A \cdot U_1 \cdot U_2 \dots U_D = U_f,    
\end{equation}

from which follows

\begin{equation}
\label{decomp}
A = U_f \cdot U_D^\dag \cdot U_{D-1}^\dag \dots U_1^{\dag}.
\end{equation}

Equation \eqref{decomp} gives desired decomposition of $A$ into 2-level unitary matrices. This decomposition has $\frac{d(d-1)}{2}$ matrices. Indeed, each matrix in decomposition (including $U_f$) corresponds to one eliminated element in upper triangular part of matrix $A$, and there are $\frac{d(d-1)}{2}$ such elements.

All matrices in decomposition are special unitary, with two exceptions. First, if we were swapping columns due to $a$ being zero, we will have two-level $X$ matrices. Second, if $\det(A) \ne 1$, matrix $U_f$ will not be special unitary.

\subsection{Gray codes}

Now all two-level matrices in decomposition act on pair of states $(i, i+1)$.
For our purposes we want them to act on pairs of states differing only in one bit, i.e. $(i, i \oplus 2^k)$. 

Luckily, for any positive integer $n$ exists such permutation of numbers $0, 1, \dots, 2^n-1$, that any two neighboring numbers in it differ only in one bit. Such permutation is called binary-reflected Gray code \cite{gray1953pulse}, and is given by formula

\begin{equation}\pi_i = i \oplus \left\lfloor i/2 \right\rfloor, 
\end{equation}
where $i=0, 1, \dots, 2^n-1$.

For example, Gray code for $n=3$ is $(0, 1, 3, 2, 6, 7, 5, 4)$.

Let's consider matrix $P \in \mathrm{U}(2^n)$, such that 
$P_{ij} = \delta_{i, \pi_j}$. This is permutation matrix, i.e. its action on a vector is permuting elements of that vector with permutation $\pi$. Then expression $P^\dag A' P$ simultaneously permutes rows and columns of matrix $A'$ with permutation $\pi$. If $A'$ was a two-level matrix acting on states $(i,i+1)$, then 
$A = P^\dag A' P$ will be two-level matrix acting on states $(\pi_i, \pi_{i+1})$ --- exactly what we need.

So, we need to apply two-level decomposition algorithm to matrix $A' = P A P^\dag$ and get decomposition 
$A' = \prod_{i}A'_i$. Then $A = P^\dag A' P = 
\prod_{i}(P^\dag A'_i P)$. So, we have decomposition of $A$ into two-level unitary matrices acting on states differing in one bit.

Similar technique is used in \cite{vartiainen2004efficient}.

\subsection{Fully controlled gates}

Let's call gate a fully controlled (FC) gate acting on qubit $i$ if this gate acts on qubit $i$ and is controlled by all other qubits in the register. This gate will act on certain basis state only if all bits in index of this state (except maybe $i$-th) are set to one. 
For example, if $n=5$, FC gate $U$ acting on bit $1$ applies matrix $U$ to states $\ket{10111}$ and $\ket{11111}$.

By convention, FC gate acting on single qubit is just simple one-qubit gate without control qubits.

FC gate applies a two-level unitary matrix. But also any two-level unitary matrix (acting on states differing in one bit) can be implemented with a fully controlled gate and possibly some single-qubit $X$ gates. Let's show how.

Let $U$ be two-level unitary acting matrix on states $(i, i \oplus 2^r$).  Let $J_0$ --- set of all indices $j$, such that $j$-th bit of $i$ is zero, $J_1$ --- set of all indices $j$, such that $j$-th bit of $i$ is one (both sets don't include $r$). Then we need to apply this two-level unitary only to pair of such states, whose indices have zeroes on positions $J_0$, and ones on positions $J_1$. If $J_0 = \emptyset$, this is simply fully-controlled gate on qubit $r$. 

But if there is some $j \in J_0$, then we have to just apply $X$ on $j$-th qubit, then apply $U$ and then apply $X$  on $j$-th  qubit again.

This will work because $X$ acting on $j$-th qubit swaps state $i$ with state $i \oplus 2^j$, so if $U$ does something only with states $i$ where $i[j]=1$, then 
$XUX$ does the same thing with states $i$ where  $i[j]=0$.

So, to implement two-level unitary matrix we have to apply $X$ gate to all qubits from $J_0$, then apply fully-controlled gate on qubit $r$ and then again apply $X$ gate to all qubits from $J_0$.

Let's consider an example. Let $n=5$ and we want to apply two-level unitary matrix with non-trivial $2 \times 2$ submatrix $U$ acting on states $\ket{10100}$ and $\ket{10110}$. Then $r=3$,
$J_0 = \{1,4\}$ and $J_1 = \{0,2\}$. So, we need to build the following circuit:
 
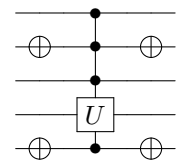
\[
\Qcircuit @C=.5em @R=0em @!R {
& \qw      & \qw & \ctrl{1} & \qw & \qw  & \qw\\
& \targ & \qw & \ctrl{1} & \qw & \targ & \qw\\
& \qw      & \qw & \ctrl{1} & \qw & \qw & \qw\\
& \qw      & \qw & \gate{U} & \qw & \qw & \qw\\
& \targ & \qw &  \ctrl{-1}  & \qw & \targ & \qw\\
}
\]

\subsection{Implementing a single gate}

At this point we have sequence of $X$ gates and FC gates acting on single qubit, but each such gate is represented by arbitrary $U \in \mathrm{U}(2)$. We have to decompose $U$ into product of matrices, which can be implemented by gates. In this paper I will consider how to decompose them into $R_1$, $R_y$ and $R_z$ gates, which are defined as:

\begin{equation}
R_1(\alpha) = \begin{pmatrix}
    1  & 0 \\
    0  & e^{i\alpha}
\end{pmatrix},
\end{equation}

\begin{equation}
R_y(\alpha) =\exp \left(\frac{i \alpha \sigma_y}{2} \right)=
\begin{pmatrix}
    \cos(\alpha/2)  & \sin(\alpha/2) \\
    -\sin(\alpha/2) & \cos(\alpha/2) 
\end{pmatrix},
\end{equation}

\begin{equation}
R_z(\alpha) =\exp \left(\frac{i \alpha \sigma_z}{2} \right)=
\begin{pmatrix}
    e^{i \alpha/2}  & 0 \\
    0 & e^{-i \alpha/2} 
\end{pmatrix}.
\end{equation}

First step is to make $U$ special unitary matrix, if it's not such matrix already.
Let $\phi = \arg{\det U}$ (recall that $|\det U|=1$).
Then $\det(R_1(-\phi) \cdot U) = e^{-i \phi} e^{i \phi} =1.$
So, takes place decomposition $U = R_1(\phi) U'$, where 
\begin{equation}
U' = R_1(- \phi) U
\end{equation}
and $U'$ is special unitary.

Now all is left is to decompose special unitary matrix $U'$ into gates.
As $U'$ is special unitary, it  can be written in form \cite[\S 4.6]{nakahara2008quantum}:

\begin{equation}
U' = \begin{pmatrix}
    \cos{\theta}e^{i \lambda}  & \sin{\theta}e^{i \mu} \\
    -\sin{\theta}e^{i \mu}  & \cos{\theta}e^{-i \lambda}
\end{pmatrix},
\end{equation}

where
\begin{equation}
\label{par_def}
\theta = \arccos(|U'_{00}|), \lambda = \arg(U'_{00}), \mu = \arg(U'_{01}).
\end{equation}

It can be directly checked that then

\begin{equation}
U' = R_z(\lambda+\mu) R_y(2 \theta) R_z(\lambda - \mu).
\end{equation}

So, action of single-qubit gate $U$ can be implemented using four gates:

\begin{equation}
U = R_1(\phi)R_z(\lambda+\mu) R_y(2 \theta) R_z(\lambda - \mu),
\end{equation}

where $\phi = \arg{\det U}$ and other parameters are given by \eqref{par_def}.
\subsection{Optimizations}

Combining all previous steps, we can build sequence of single-qubit $X$ gates and FC gates $R_1$, $R_y$ and $R_z$ which implements given unitary matrix. 

Each of $\mathcal{O}(4^n)$ FC gates is surrounded by $O(n)$ X-gates. It can happen that there are two $X$ gates (after one FC gate and before next FC gate), which act on the same qubit. As $X^2=I$, they both can be removed.

This will eliminate significant amount of $X$ gates. When neighboring two-level matrices correspond to eliminating elements of the same row, they act on states $(i_1, i_2)$ and $(i_2, i_3)$, so mask $J_0$ for them can differ at most in two bits, so there will be not more than two $X$ gates between almost all FC gates after optimization. There is only $\mathcal{O}(2^n)$ pairs of gates where this doesn't work and we can have up to $n$ X gates (this happens when we proceed to next row in matrix). Overall, this guarantees that after optimization there will be $\mathcal{O}(n \cdot 2^n + 4^n) = \mathcal{O}(4^n)$ X gates, or $\mathcal{O}(1)$ X gates per one two-level matrix.

Another easy optimization to make is to remove all gates which are identity matrices, namely $R_1(2\pi k), R_y(4 \pi k), R_z(4 \pi k)$ for $k \in \mathbb{Z}$.

One more optimization is when we convert matrix to gates and matrix $X$ occurs, don't apply usual procedure for not-special unitary matrix (which will result in decomposition $X = R_y(-\pi) R_1(\pi)$), but just use FC gate X instead. If we do that, we can guarantee that final circuit will contain at most one $R_1$ gate, and even this gate will be needed only if initial matrix was not special unitary.

\subsection{Implementation}

I implemented a Python program which uses described algorithm to transform arbitrary uniform matrix $U \in \mathrm{U}(2^n)$ into a Q\# operation, which implements action of this matrix on array of $n$ qubits. This program is available on GitHub \cite{quantumDecompGithub}. 

This program performs all steps described above and then maps gates to standard  Q\# commands, namely \texttt{X, Controlled X, Controlled Ry, Controlled Rz, Controlled R1}. 

\section{Complexity}

One interesting question to consider is how many operations does it generally require to implement unitary matrix acting on $n$ qubits.

Decomposition into two-level unitary matrices consisted of $\frac{2^n(2^n-1)}{2} = \mathcal{O}(4^n)$ matrices. Each two-level matrix was mapped to 3 or 4 fully-controlled gates and $\mathcal{O}(n)$ X gates, but after optimization we expect to have $\mathcal{O}(1)$ X gates per two-level unitary matrix. Thus we expect number of needed commands to be $\mathcal{O}(4^n)$.

\bigskip

Let's check it experimentally. I generated random matrices $U \in \mathrm{U}(2^n)$ for $n=1\dots 9$, decomposed them using described algorithm, and calculated how many gates of each type appears in the decomposition ($\#(X), \#(R_y), \#(R_z), \#(R_1)$). Also I calculated total number of gates $G(n)$ and number of gates per matrix element, which is number of gates divided by $4^n$. Results are shown in table \ref{table:experiment}. 

As we can see from the data, indeed $G(n) \sim 2.00 \cdot 4^n = \mathcal{O}(4^n)$.

\begin{table}
\caption{Number of gates in decomposition of random matrices}
\label{table:experiment}
\begin{tabular}{|l|l|l|l|l|l|l|l|l|}
\hline
n & \#($X$)    & \#($R_y$)    & \#($R_z$)    & \#($R_1$)    & $G(n)$    & $G(n)/4^n$   \\ \hline
1 & 0    & 1    & 2  & 1    & 4    & 1.00  \\ \hline
2 & 2    & 6    & 12  & 1  & 21    & 1.31  \\ \hline
3 & 28    & 28    & 56  & 1  & 113    & 1.77  \\ \hline
4 & 130    & 120    & 240  & 1  & 491    & 1.92  \\ \hline
5 & 532    & 496    & 992  & 1  & 2021    & 1.97  \\ \hline
6 & 2118    & 2016    & 4032  & 1 & 8167  & 1.99  \\ \hline
7 & 8392    & 8128    & 16256  & 1 & 32777   & 2.00 \\ \hline
8 & 33290    & 32640    & 65280  & 1 & 131211 & 2.00  \\ \hline
9 & 132364    & 130816    & 261632  & 1 & 524813 & 2.00  \\ \hline
\end{tabular} 
\end{table}

\bigskip
Let's denote $G_{min}(n)$ --- minimal number of gates ($X$,  FC-$R_y(\alpha)$, FC-$R_z(\alpha)$, FC-$R_1(\alpha)$), needed to implement arbitrary  $2^n \times 2^n$ unitary matrix. We showed that $G_{min}(n) = \mathcal{O}(4^n)$.
 
Unitary matrix $2^n \times 2^n$ can be parametrized by $4^n$ independent real numbers \cite[\S IV.4]{zee2016group}. So, to implement arbitrary matrix using gates parametrized by a real number, we will need at least $4^n$ such gates. So, we will have to use at least $4^n$ FC $R_1, R_y$ and $R_z$ gates, which gives us lower bound $G_{min}(n) = \Omega(4^n)$. Therefore, proposed algorithm gives asymptotically optimal result.

However, in some special cases matrix can be represented by much fewer number of gates, for example, if it was built as product of a few single-qubit gates and CNOT gates. Generally, proposed algorithm will not recognize special structure of the matrix and will still return $\Omega(4^n)$ gates.

\bigskip
\bigskip


\section{Discussion}

\subsection{Further decomposition}

In this paper we consider fully controlled $R_y$, $R_z$ and $R_1$ gates as primitives, because they are supported by Q\# language. However, two-level matrix can be decomposed into sequence of simpler gates, namely $R_y$, $R_z$ and $R_1$ gates acting on single qubit and CNOT (controlled-X) gate (acting on two qubits), and decomposition of a single two-level matrix will contain $\Theta(n^2)$ such gates. How to do that is shown in \cite[\S 4.6]{nakahara2008quantum}.

\subsection{Universality}

Set of gates is \textit{universal} if any quantum gate acting on any number of qubits can be implemented by combining gates from this set.

Universality is important property for quantum computers: if we have set of universal gates, we only need to be able to implement them on a particular quantum computer to be able to implement any irreversible computation on it. Possibility of implementing universal set of gates is one of DiVincenzo criteria \cite[\S 11]{nakahara2008quantum}, meaning it is necessary requirement for any physical system to be a viable quantum computer.

This paper can be seen as a constructive proof that set of single-qubit X gate and fully-controlled gates $R_x(\alpha), R_y(\alpha), R_1(\alpha)$ with any number of control bits (including 0) and with any parameters is an universal set of gates.

However, this set can be reduced to just set of CNOT gate and single-qubit gates $R_y(\alpha), R_z(\alpha), R_1(\alpha)$ \cite[\S 4.6]{nakahara2008quantum}. Moreover, this set can be reduced to just one 2-qubit gate \cite{deutsch1995universality}.


\bibliographystyle{utphys}    
\bibliography{references}     

\end{document}